\documentclass{an}
\usepackage{times}
\usepackage{latexsym,graphicx,rotating,amsmath, epsfig}
\usepackage{natbib}
\def\sol{\odot}

\begin{document}

\Pagespan{1002}{}

\Yearpublication{2007}%
\Yearsubmission{2007}%
\Month{11}%
\Volume{328}%
\Issue{10}%
\DOI{10.1002/asna.200710843}%

  \title{Rapid Rotation, Active Nests of Convection and Global-scale
  Flows in Solar-like Stars}
  \author{B.P.\ Brown \inst{1}\fnmsep\thanks{Corresponding author:
  \email{bpbrown@solarz.colorado.edu}}
 \and M.K.\ Browning \inst{2}
  \and A.S.\ Brun \inst{1,3} \and M.S.\ Miesch \inst{4} \and J.\ Toomre \inst{1}} 
  \institute{JILA and Dept.\ of Astrophysical \& Planetary Sciences,
    University of Colorado, Boulder, CO 80309-0440 \and 
    Dept.\ of Astronomy, University of California, Berkeley, CA
    94720-3411 \and 
    DSM/DAPNIA/SAp, CEA Saclay, Gif sur Yvette, 91191 Cedex, France\and
    High Altitude Observatory, NCAR, Boulder, CO 80307-3000}

  \publonline{at http://www.aip.de/AN}


  \abstract{%
    In the solar convection zone, rotation couples with intensely
    turbulent convection to build global-scale flows of differential
    rotation and meridional circulation.  Our sun must
    have rotated more rapidly in its past, as is suggested by
    observations of many rapidly rotating young solar-type stars.
    Here we explore the effects of more rapid rotation on the patterns
    of convection in such stars and the global-scale flows which are self-consistently
    established.  The convection in these systems is richly
    time dependent and in our most rapidly rotating suns a striking
    pattern of spatially localized convection emerges.  Convection near the
    equator in these systems is dominated by one or two patches of
    locally enhanced convection, with nearly quiescent streaming flow
    in between at the highest rotation rates.  These active nests of
    convection maintain a strong differential rotation despite their
    small size.  The structure of differential rotation is similar in
    all of our more rapidly rotating suns, with fast equators and
    slower poles.  We find that the total shear in differential
    rotation, as measured by latitudinal angular velocity contrast,
    $\Delta \Omega$, increases with more rapid rotation while the relative
    shear, $\Delta \Omega/ \Omega$, decreases.  In contrast, at more
    rapid rotation the meridional circulations decrease in both energy
    and peak velocities and break into multiple cells of circulation
    in both radius and latitude.}

\maketitle

\section{Rotation and Dynamo Action}
Rotation and magnetism couple with turbulent plasma motions in stellar
convection zones to drive dynamos and cycles of magnetic activity.
When our sun was younger, it must have rotated more rapidly, as is
suggested by observations of rapidly rotating young solar-like stars
and by the presence of the magnetized solar wind, which continually
removes angular momentum from the star.  Observations of rapidly
rotating stars indicate that generally more rapid rotation is
correlated with stronger magnetism and perhaps stronger dynamo action.
In the sun, global-scale dynamo action is thought to arise from the
coupling of convection and rotation.  Here we explore the effects of
more rapid rotation on global-scale convective structures and in
particular on the resulting differential rotation and meridional
circulation.

Helioseismology, which uses acoustic oscillations at the solar surface
to probe the radial and latitudinal structure of the star as well as the convective
flows beneath the surface, has revealed that the solar differential
rotation profile observed at the surface prints throughout the
convection zone with two prominent regions of radial shear.  The
near-surface shear layer exists in the outer 5\% of the sun, and the
tachocline lies between the differentially-rotating convection zone
and the deeper radiative interior, which is in nearly solid body
rotation \citep{Thompson_et_al_2003}.  These shear layers, combined
with the global-scale flows of meridional circulation and differential
rotation, are thought to be the key components in the global solar
dynamo which builds and rebuilds magnetic fields yielding 22-year
cycles of magnetic activity.  In the interface dynamo model
\citep[e.g.,][]{Ossendrijver_2003,Charbonneau_2005}, magnetic fields generated in the
bulk of the convection zone are pumped downward into the tachocline where the
strong radial shear builds large-scale fields that eventually erupt at
the solar surface.  The differential rotation plays an important role
both in building and organizing the global-scale fields while the
meridional circulations may be important for returning flux to the
base of the convection zone, enabling cycles of magnetic activity.
The nature of internal differential rotation in other stars is
unclear, though some observations of the surface differential rotation
are now possible through a variety of techniques including photometric
variability \citep{Donahue_et_al_1996, Walker_et_al_2007}, Doppler
imaging \citep{Donati_et_al_2003} and Fourier transform methods
\citep{Reiners&Schmitt_2003}.  We undertake simulations to
self-consistently establish differential rotation through the
coupling of convection and rotation to study this crucial ingredient
for stellar dynamo action.

\section{Simulations of rapidly rotating suns}
\label{sec:ASH}
To capture the essential couplings between convection, rotation and
the resulting global-scale flows, we must employ a global model which 
simultaneously includes the spherical shell geometry and admits the
possibility of zonal jets, large eddy vortices and convective plumes
which may span the depth of the convection zone.  The solar convection
zone is intensely turbulent and molecular values of viscosity and
thermal diffusivity in the sun are estimated to be very small.  As a
consequence, numerical simulations cannot hope to resolve all scales
of motion present in real solar convection and a compromise must be
struck between faithfully capturing the important dynamics within
small regions and capturing the connectivity and geometry of the
global scales.  

\subsection{The ASH code}
Our tool for exploring stellar convection is the anelastic spherical
harmonic (ASH) code, which is described in detail in
\cite{Clune_et_al_1999} and in \cite{Brun_et_al_2004}.  ASH is a
mature simulation code, designed to run on massively parallel
architectures, which solves the three-dimensional MHD equations of
motion under the anelastic approximation.  The velocity field is fully
nonlinear, but under the anelastic approximation the thermodynamic
variables are linearized about their spherically symmetric and
evolving mean state with density $\bar{\rho}$, pressure $\bar{P}$,
temperature $\bar{T}$ and specific entropy $\bar{S}$ all varying with
radius.  The overall density contrast is about 25 between the top and
bottom of the simulation.

As a global-scale code, ASH cannot resolve all scales of motion present
in real stellar convection zones.  Instead, we conduct large eddy
simulations (LES) which explicitly resolve the largest scales and
describe scales below the grid resolution with subgrid-scale (SGS)
modelling.  Here we treat these scales with effective eddy diffusivities
$\nu$~and~$\kappa$, which represent momentum and heat transport by
unresolved motions in the simulations. For simplicity
$\nu$~and~$\kappa$ 
are taken as functions of radius alone and in the simulations
reported are proportional to $\bar{\rho}^{-1/2}$.  The Prandtl
number $Pr = \nu/\kappa$ is 0.25 for all simulations.

\subsection{Our simulations}
We are interested here primarily in the interaction between rotation
and convection in deep stellar convection zones.
In these simulations, we avoid the H and He ionization regions
near the stellar surface as well as the stably stratified
radiative interior and the tachocline of shear at the base of the convection
zone.  Thus we are concerned principally with
the bulk of the convection zone and our computational
domain extends from $0.72R_{\sol}$ to $0.96R_{\sol}$.  
Solar values are taken for heat flux, mass and radius, and a perfect
gas is assumed.   The reference
or mean state of our thermodynamic variables is derived from a
one-dimensional solar structure model \citep{Brun_et_al_2002} and is
continuously updated with the spherically-symmetric components of the
thermodynamic fluctuations as the simulations proceed.  
We have conducted a number of 3-D hydrodynamic simulations for stars
rotating from one to ten times the current solar rate.
Here we concentrate on cases at the solar rate (case~G1) and at five times the
solar rate (G5); the full suite of simulations are discussed at length in
\cite{Brown_et_al_2007}.  The parameter space explored by these
simulations is presented in Table~\ref{table:sim parameters}.  As we
probe higher rotation rates, we decrease the eddy
diffusivities in the simulation to maintain a comparable level of
supercriticality.

\begin{table}
\centering
\caption{Simulation parameter space.  
	Evaluated at mid-depth are the
	Rayleigh number $R_a$,
	the Taylor number $T_a$,
	the rms Reynolds number $R_{e}$ and
	fluctuating Reynolds number $R_{e}'$,
	the Rossby number $R_\mathrm{o}$
	and convective Rossby number $R_\mathrm{oc}$,
	and the diffusivities $\nu$ and $\kappa$ in $\mathrm{cm}^{2}/\mathrm{s}$.
        The rotation rate of the reference frame $\Omega_0$ is given in multiples
        of $\Omega_\sol=2.6 \times 10^{-6}$ rad/s or $414$ nHz.
	Also quoted are kinetic
	energies relative to the rotating system
	contained in convection (CKE), differential rotation (DRKE)
	and meridional circulations (MCKE) in units of $10^6$ erg
	cm$^{-3}$ and as a percentage of the total.}
\label{table:sim parameters}
\begin{tabular}{ccc}\hline
& & \\[-7pt]
case & G1 & G5 \\
\hline
& & \\[-7pt]
$\Omega_0$ & $1~\Omega_\odot$ & $5~\Omega_\odot$ \\
$R_a$ &  
$ 3.22 \times 10^{4} $ & 
$ 1.29\times 10^{6} $ \\
$T_a$ &
$ 3.14 \times 10^{5} $ &
$ 6.70\times 10^{7} $ \\ 
$R_e$ &  
$ 84$ & 
$ 543$ \\ 
$R_e'$ &  
$ 63$ & 
$ 133$ \\ 
$R_\mathrm{o}$ &
$  0.92$ & 
$  0.28$ \\ 
$R_\mathrm{oc}$ &
$  0.61$ & 
$  0.27$ \\ 
$\nu$ &
$  2.75 \times 10^{12}$ & 
$  0.94 \times 10^{12} $ \\ 
$\kappa$ &	      
$ 11.0 \times 10^{12} $ &
$ 3.76 \times 10^{12} $ \\
& & \\[-7pt]
CKE  & $   3.28$ ( $   59.0\% $) & $   2.11$ ( $   6.55\% $) \\
DRKE & $   2.26$ ( $   40.6\% $) & $   30.1$ ( $   93.4\% $) \\ 
MCKE & $ 0.025$ ( $  0.45\% $) & $0.007$ ( $ 0.020\% $) \\

\hline
\end{tabular}
\end{table}

\begin{figure*}[t]
  \includegraphics{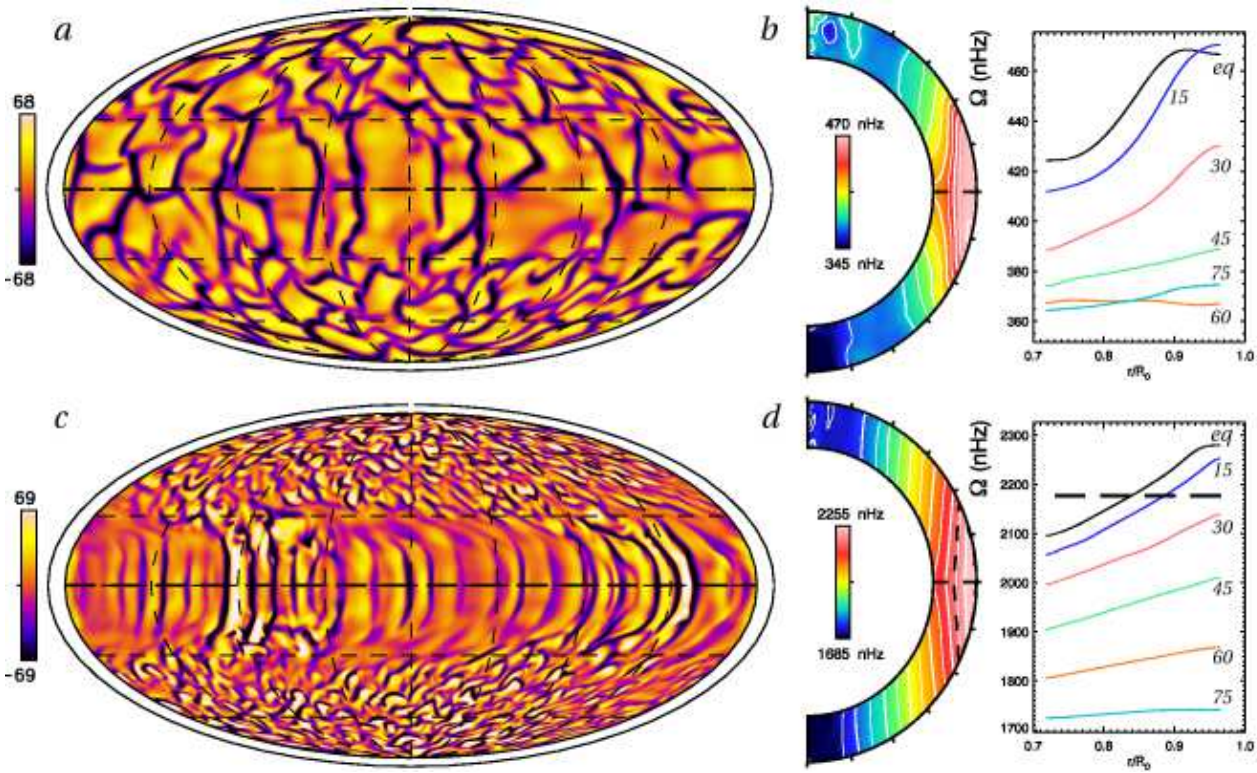}

  \caption{Evolution of global-scale convective flows with increasing
  rotation rate for case~G1~(\emph{a,b}) and case~G5~(\emph{c,d}).
  Shown in $(a,c)$ are global views, in
  Mollweide projection, of radial velocity near the stellar surface
  ($r=0.95 R_\sol$).
  Upflows are light while downflows are dark, with scales indicated in
  m/s.  At high rotation rates a striking pattern
  of modulated convection emerges at low latitudes, consisting of
  spatially modulated or patchy convection.  Shown in $(b,d)$ 
  are azimuthal averages of angular velocity $\Omega$ with radius and latitude.
  These have been further averaged in time over a period of roughly
  200 days.  Plotted at right
  are radial cuts of angular velocity at selected latitudes as
  indicated.  The black dashed contour in (\emph{d}) denotes the constant
  propagation rate of the nests in such modulated convection.
  \label{fig:ab2_turf}}
\end{figure*}

\section{Convection in more rapidly rotating suns}
Examples of convective patterns in case~G1 and G5 are
illustrated in Figure~\ref{fig:ab2_turf}.  The radial
velocity near the top of the domain is presented in Mollweide
projection, which displays the entire layer with
minimal distortion, with poles at top and bottom and the entire
equatorial region displayed in the middle.   
The convection patterns are complex, time dependent and asymmetric.
Owing to the density stratification, the narrow, fast downflow lanes are
surrounded by broad, relatively weak upflows.  Convection in the
equatorial regions is dominated by large cells aligned with the
rotation axis whereas at high latitudes the convection is more isotropic
and cyclonic.  With more rapid rotation, the convection cells decrease
in horizontal size and align more strongly with the rotation axis 
but remain vigorous and turbulent.

A striking finding in our more rapidly rotating systems is the pattern
of spatially localized convection evident in
Figure~\ref{fig:ab2_turf}$c$.  This structure is a region of locally
enhanced convection spanning from about $\pm 30^\circ$ in latitude.  
Here convective velocities and transport are enhanced, 
and strong plumes span the depth of the convection zone.  
This modulated pattern emerges
gradually with more rapid rotation and we have found states with
either multiple patches of enhanced convection or single isolated
patches.  At the highest rotation rates, convection in the equatorial
regions is present only within the patches, with complex streaming
zonal flows in the interpatch regions. 

These active nests of convection persist for long periods of time
(thousands of days) compared to the typical lifetime of individual
convective cells ($\sim10-40$ days).  They propagate in a prograde fashion
relative to the stellar rotation rate and with  
a fixed angular velocity at all depths and latitudes in the
equatorial region.  The nests of convection are embedded in a strong
radial shear, as shown in Figure~\ref{fig:ab2_turf}$d$.  Near the
surface the mean zonal flow streams through the nests while at depth
the zonal flows are slower than the nests.  The strong radial shear in the rapidly
rotating cases leads to convective structures which are typically
double-celled in radius.  Near the stellar surface the individual convective cells
propagate more rapidly than the nests, with these cells catching up
and swimming through the structure.  Near the
base of the convection zone the nests propagate faster than
the individual convective cells.

\begin{figure*}[tb]
  \includegraphics{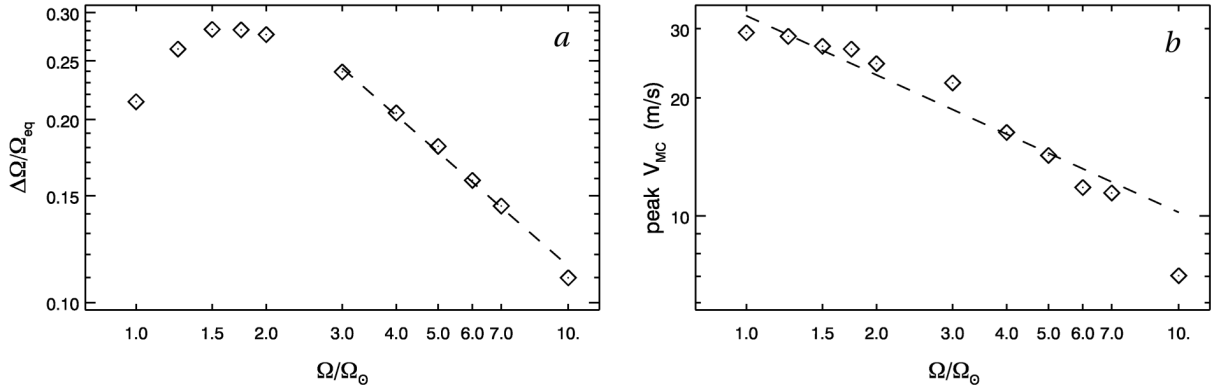}
\caption{Evolution of circulations with faster rotation: 
  $(a)$ angular velocity contrast $\Delta \Omega/\Omega_\textrm{eq}$
  plotted against bulk rotation rate $\Omega/\Omega_\odot$  of the simulations
  in logarithmic scaling. 
  A power law with exponent $n=-0.6$ is overplotted.
  $(b)$ Peak meridional circulations at the top of the simulation
  ($r=0.96 R_\sol$), showing a steep decline
  with more rapid rotation as the circulations break into multiple
  cells aligned with the rotation axis.
  Here a power law with exponent $n=-0.5$ overlies the data.
  \label{fig:ratios_of_omega}}
\vspace{0.25cm}
\end{figure*}

\section{Mean flows and circulations}
Differential rotation in our simulations is primarily established
by Reynolds stresses from the convection.  Convective
enthalpy transport in latitude establishes a strong latitudinal
temperature contrast in the more rapidly rotating stars, which are
largely in thermal wind balance with their differential rotation.  The
profile of angular velocity realized in case G1
(Fig.~\ref{fig:ab2_turf}$b$) is similar to the solar internal rotation
profile, with a fast equator and a monotonic decrease in angular
velocity towards the poles.  The angular velocity profile realized in
this simulation is significantly more columnar than the
helioseismically deduced solar
internal rotation profile, which is nearly constant on radial lines.
Recent simulations \citep{Miesch_et_al_2006} reveal that a more
solar-like rotation profile can be achieved by including a latitude
dependent thermal forcing that is consistent with a tachocline in
thermal wind balance.  As the nature of tachoclines and the magnitude
of this thermal forcing is unclear in other stars, we have chosen here
to neglect this latitude dependent thermal forcing and thus our bottom
thermal boundary condition is one of constant flux.
All of our rapidly rotating stars build angular velocity profiles with
fast equators and slow poles.  This strong differential rotation is
realized even in the most rapidly rotating cases, which exhibit more
extreme degrees of convective modulation.

The kinetic energy contained in the differential rotation (DRKE) grows
with rotation rate (Table~\ref{table:sim parameters}).
This is accompanied by a growth in the total latitudinal angular
velocity contrast $\Delta\Omega=\Omega_\mathrm{eq}-\Omega_{60^\circ}$.  
This contrast grows more slowly than linearly with increasing rotation
rate, and thus the relative contrast $\Delta \Omega/\Omega$
decreases in the rapidly rotating simulations
(Fig.~\ref{fig:ratios_of_omega}$a$).  Our simulations of rapidly
rotating suns appear to roughly follow a power law scaling at the highest
rotation rates.  However, the actual curve and scaling depends on the
parameter space chosen for the simulations, and thus on the level of
turbulence achieved, as discussed in detail in \cite{Brown_et_al_2007}.

The energy contained in meridional circulations (MCKE) declines in both magnitude and
percentage with more rapid rotation. The decrease in MCKE is
accompanied by a decrease in the magnitude of meridional velocities at
the surface, as shown in Figure~\ref{fig:ratios_of_omega}$b$, where
we plot the magnitude of the peak time-averaged meridional velocity at the
surface.  This decrease in circulation strength is
accompanied by a change in the topology of the circulations.  In case
G1, the meridional circulations are largely single-celled in radius
and latitude in each hemisphere.  In the more rapidly rotating stars,
as in case G5, these cells break into multiple weak cells in both
radius and latitude, and near the equator these cells align with the
rotation axis.

\section{Conclusions}
The complex coupling between rotation and convection drives a strong
differential rotation in these more rapidly rotating stars.  The
meridional circulations in contrast are much weaker, decreasing in
both magnitude and relative contribution with more rapid rotation.  As
such, more rapidly rotating stars may have dynamos which differ
substantially from the solar dynamo, where meridional circulations are
thought to play an important role.  These mean circulations are
achieved despite the presence of strong spatial modulation in the
convection of the equatorial regions.  If these active nests of
convection persist in stellar convection zones in the presence of
magnetism, they may lead to features at the surface which propagate at
rates distinct from that of the surface convection or the zonal flows
of differential rotation.  Such strong and persistent modulation in
convective structures may have important observational consequences.

\acknowledgements
This research was supported by NASA through Heliophysics Theory
Program grant NNG05G124G and the NASA GSRP program by award number
NNG05GN08H.  The simulations were carried out with NSF PACI support of
PSC and SDSC.

\bibliographystyle{apj}
\def\aap{A\&A}
\def\aapr{A\&A Rev.}
\bibliography{bibliography}

\end{document}